\documentclass[10pt,twocolumn,english,prl,superscriptaddress]{revtex4}
\usepackage{times}
\usepackage{mathptm}
\usepackage[T1]{fontenc}
\usepackage[latin1]{inputenc}
\usepackage{graphicx,epsfig}

\begin{document}
\voffset=0.7in
\def\deg{^\circ}
\def\AA{{\rm A}}
\def\half{ \small {1 \over 2}}
\bibliographystyle{./simon}

\voffset=0.7in
\def\deg{^\circ}
\def\AA{{\rm A}}
\def\half{ \small {1 \over 2}}
\bibliographystyle{./simon}

\title{How a liquid becomes a glass both on cooling and on heating}

\author { Xinhui Lu}
\affiliation{Department of Physics, Yale University, New Haven, Connecticut 06511}
\author{S. G. J. Mochrie}
\affiliation{Department of Physics, Yale University, New Haven, Connecticut 06511}
\affiliation{Department of Applied Physics, Yale University, New Haven, Connecticut 06511}
\author{S. Narayanan}
\affiliation{Advanced Photon Source, Argonne National Laboratory, Argonne, IL 60439}
\author{A. R. Sandy}
\affiliation{Advanced Photon Source, Argonne National Laboratory, Argonne, IL 60439}
\author{M.  Sprung}
\affiliation{Advanced Photon Source, Argonne National Laboratory, Argonne, IL 60439}

\date{\today}


\begin{abstract}
The onset of structural arrest and glass formation
in a concentrated suspension of
silica nanoparticles in a  water-lutidine binary mixture near its consolute
point is studied by exploiting the near-critical fluid degrees of freedom
to control the strength of an attraction between particles and
multispeckle x-ray photon correlation spectroscopy to determine the particles'
collective  dynamics.  This model system undergoes a glass transition
both on cooling and on heating, and the intermediate liquid realizes unusual
logarithmic relaxations.  How vitrification occurs for the two different glass
transitions is characterized in detail and comparisons are drawn to
theoretical predictions for glass formation in systems with attractive interactions.
\end{abstract}

\maketitle

%Understanding
The glass transition remains
a grand challenge for condensed matter
science \cite{Angell2001Nature,Berthier2005},
with applications in diverse areas such as protein folding \cite{Brooks2001}
and the flow of granular
materials \cite{LiuNature1998}.
Especially remarkable then
are
predictions \cite{PhysRevE.59.5706,PhysRevE.59.R1347,DawsonPRE2001,PhysRevE.66.041402},
that in systems with short-ranged attractions
a dense liquid may become a glass both on
cooling and on heating.
The scenario is as follows:
When the ratio of attraction strength to temperature is small,
%At high temperatures,
the attractions play no role and, at high densities,
a repulsion-dominated glass (RG) is realized
 \cite{van_megan:91,gotze:91}. However, as the ratio
is increased, surprisingly, attractions cause the glass to
melt. With still further increase,
the resultant liquid
re-vitrifies into an attraction-dominated glass (AG).
The
intermediate
liquid
is itself highly unusual in that its density fluctuations
are predicted to relax logarithmically in time.
In the present paper,
by exploiting
fluid-mediated interactions
between silica particles
in a near-critical water-lutidine binary mixture
to control the attraction
strength \cite{BeysensPRL1985,SchlesenerJSP2003,PontoniPRL2003},
and
multispeckle x-ray photon correlation
spectroscopy (XPCS) \cite{dierker:95,mochrie:97,falus:04}
to determine the
collective  dynamics,
we show that
re-entrant glassy behavior
and logarithmic relaxations are realized in this  model system,
simply by changing temperature.
We are thus able to characterize in unprecedented detail
how the  non-ergodicity parameter of the
RG decreases upon attraction-driven
melting, and how relaxations in
the intermediate fluid arrest as the attactive glass phase is approached.

%Recent experiments on colloidal suspensions have shown that
Glassy states in colloidal suspensions
are essentially spatially and temporally inhomogeneous
 \cite{WeeksScience,KegelScience}. It is
therefore paradoxical that mode-coupling theory (MCT)
-- a mean-field-like approach --
is the
only theory to-date  capable of quantitative predictions
concerning collective dynamics
near the glass transition \cite{gotze:91}.
Nevertheless, MCT captures essential aspects
of the dynamical arrest observed near the
prototypical glass transition that occurs in a
suspension of
hard spheres  \cite{van_megan:91,gotze:91},
where
%In this case,
arrest occurs at sufficiently high densities, when each
particle becomes trapped in a cage
% created by
of its neighbours.

Instead of caging, it is intuitive that
sufficiently strong interparticle attractions permit an alternate route
to dynamical arrest, in which particles simply stick together.
Recent
MCT calculations carried out to investigate this possibility
lead to the novel
predictions of
Refs. \onlinecite{PhysRevE.59.5706,PhysRevE.59.R1347,DawsonPRE2001,PhysRevE.66.041402}.
%Thus,
%particles with  a hard-core repulsion,
%supplemented by a short-ranged attraction,
%represent an important new arena in which to
%investigate glassy behavior and structural arrest experimentally.
Support for
these predictions
% of Refs.~
%\onlinecite{PhysRevE.59.5706,PhysRevE.59.R1347,DawsonPRE2001,PhysRevE.66.041402}
comes
from optical PCS
experiments
 \cite{PhamScience2002,EckertPRL2002,chen:03,PhamPRE2004},
in which a depletion attraction between particles was
created by the addition of non-adsorbing
polymer.
However,
how the behavior evolves
with attraction strength was only determined with limited
resolution,
in part
because  tuning the
attraction requires laboriously mixing samples.
By contrast,
silica particles in near-critical
water-lutadine (WL) mixtures
experience a temperature-dependent attraction
as a result of the confinement of the fluid  critical fluctuations
between particles
\cite{SchlesenerJSP2003},
and evidenced by the particles'
reversible aggregation as
the temperature is increased towards
the two-phase region \cite{BeysensPRL1985}.
Recently,
Pontoni et al. \cite{PontoniPRL2003}
confirmed the temperature-dependent attraction
via small-angle x-ray scattering (SAXS) and XPCS
in a dilute sample.
The present study
was carried out to explore in detail how glassy behavior
depends on attraction strength in
{\em concentrated} suspensions of silica particles in near-critical WL mixtures.

\begin{figure}
{\includegraphics[width=3in,keepaspectratio=true]{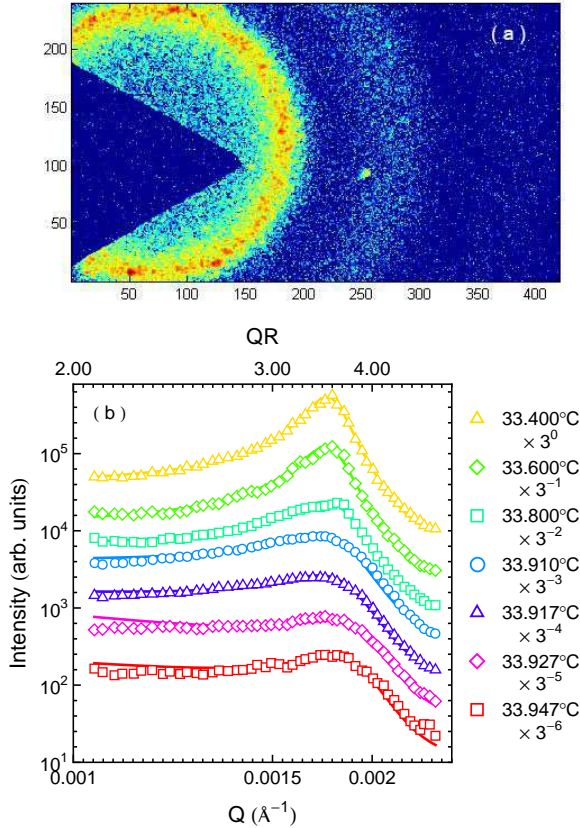}}
\caption{ (a) False-color image of the SAXS pattern from
concentrated silica spheres in WL, obtained at beamline 8-ID at the
Advanced Photon Source (APS) in Run A with a 0.15~s exposure. (b)
Average SAXS intensity vs. wave vector obtained for several
temperatures on warming (Run A). The lines are the best fits to the
model described in the text. } \label{fig1}
\end{figure}

\begin{figure}
{\includegraphics[width=3.2in,keepaspectratio=true]{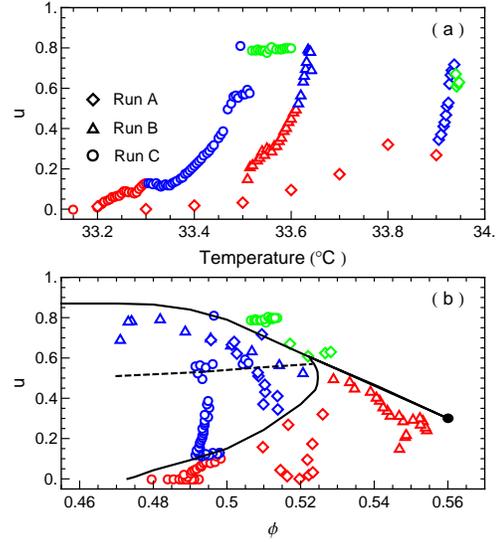}}
\caption{
(a) Best fit values of $u$ for Run A (diamonds), B (triangles) and
C (circles). The different colors indicate the different phases realized:
red corresponds to a RG, blue to a liquid,
and green to an AG.
The fitted value of $u$ appears constant in the AG,
%corresponding to an unchanging SAXS intensity and
suggesting that no structural evolution occurs in this phase.
(b) Phase diagram in the $u-\phi$ plane. Diamonds, triangles, and
circles correspond to data points from Runs A, B, and C, respectively.
Red, green, and blue symbols correspond to RG, AG, and liquid phases, respectively. Solid lines are schematic phase boundaries,
inspired by Refs.~
\onlinecite{PhysRevE.59.5706,PhysRevE.59.R1347,DawsonPRE2001,PhysRevE.66.041402},
with the black
circle corresponding to the MCT $A_3$ singularity.
The dashed line separates the region of logarithmic
relaxations above the line, from the region of stretched exponential
relaxations below.
Although our particles are subject
to gravitational settlement,
the volume fraction across the height of the x-ray beam (20~$\mu$m)
was approximately constant to within 1\%. Usually, the volume fraction
also remained constant in time. In Run B, however, the volume
fraction varied
as a result of slow settlement.
} \label{fig2}
\end{figure}

Illustrated in Fig. \ref{fig1}(a) is
the SAXS
from 195~nm-radius silica spheres (polydispersity < 10\%)
at a volume fraction of
$\phi \simeq 0.52$ in WL with a lutadine concentration of $c = 0.24$.
Partially coherent illumination gives rise to a
speckled intensity, whose mean and
time-autocorrelation characterize the
sample's static structure and dynamic relaxations, respectively.
To this end,
sequential SAXS patterns were collected
using
CCD-based area detectors at frame rates of up to  330~Hz (Run A and Run B)
or 60 Hz (Run C) \cite{falus:04}.
Fig. \ref{fig1}(b) shows
the time and azimuthally averaged SAXS intensity (symbols)  plotted versus
wavevector ($Q$)
in the neighbourhood of the peak in the interparticle structure factor,
obtained
at several temperatures during Run A.
At the lowest temperature shown (33.400$^\circ$C), there is a relatively
narrow peak at
$QR=3.6$.
With increasing temperature,
the peak gradually becomes broader
and shifts slightly toward larger
$QR$,
indicating a decrease in the
correlation length, accompanied by slight decrease in the
mean particle spacing. Concommitantly,
the intensity increases for  $QR \le  2.5$,
corresponding to the emergence of larger
scale
inhomogeneities. To quantify this evolution,
we have fitted these
data to the structure factor
of spheres subject to a potential consisting of a hard core repulsion
together with an attractive square well within the mean-spherical
approximation
\cite{DawsonPRE2001}.
The best fits
-- the solid lines in the figure -- satisfactorily
reproduce the main features of
the data, especially in view of the fact that
the only parameters varied in each fit were the overall
intensity and the strength of the attraction ($U$). The particle radius
and the
hard core radius
($R = 200~$nm)
were determined from fits to data obtained
at the lowest temperature, where the attractive
component is negligible.
Because the
model lineshape depends only on the dimensionless
combination $ u = U \Delta / (k_B T R)$
within the $Q$-range studied,
the width ($\Delta$) of the square-well attraction was fixed to be $0.01 R$.
At each temperature,
the silica volume fraction was determined via the corresponding x-ray
transmission of the sample.
Fig.~\ref{fig2}(a) shows
the best fit values of $u$ for run A (diamonds), and two additional
warming runs:   B (triangles) and C (circles).
In each case,
$u$ takes on a small value at the lowest temperatures, but
increases rapidly at a certain onset temperature.
We ascribe the different onset temperatures for the
different runs to small variations
in the lutidine concentration between runs.

\begin{figure}
{\includegraphics[width=3.4in,keepaspectratio=true]{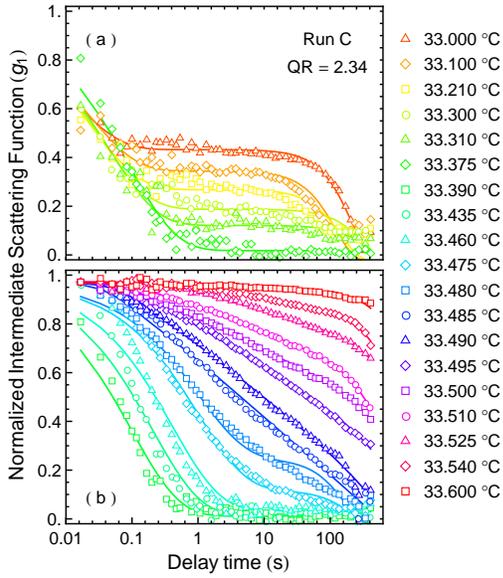}}
\caption{
$g_1$
versus delay time
%for several temperatures
during  Run C.
For
clarity, (a) corresponds
to temperatures from 33.000$^\circ$C
to 33.375$^\circ$C and
(b)
%to temperatures
from 33.390$^\circ$C
to 33.600$^\circ$C.
Points are data. Lines correspond to the models discussed in
the text.
}
\label{fig3}
\end{figure}

Normalized intermediate scattering functions ($g_1$)
\cite{falus:04} at $QR=2.34$
for delay times from 17~ms to 410~s
for a sample with $\phi \simeq 0.49$
(Run C) are illustrated
for
temperatures from 33.000$^\circ$C to 33.375$^\circ$C in
Fig.~\ref{fig3}(a),
and from 33.390$^\circ$C to 33.600$^\circ$C
in Fig.~\ref{fig3}(b).
At the lowest temperature, $g_1$ shows an
initial rapid decay (which is not fully captured by the 60~Hz frame rate of the
camera employed in Run C)
to reach a value of about 0.5 at 0.1~s.
For times
between 0.1~s and 100~s, $g_1$ remains nearly constant, revealing
arrested dynamics over three decades in time.
Such behavior is similar to that observed for
hard-sphere glasses  \cite{van_megan:91,gotze:91} and
is usually interpreted in terms of caging.
The plateau value ($f$) is called the ``non-ergodicity parameter''.
Because of the limited time range probed in the present study ($\le 410$~s), we
cannot determine whether the decrease in $g_1$ observed
for times longer than about 200~s
is a long-time equilibrium relaxation or is associated with aging of
an arrested state \cite{band:04}.
As the temperature is increased,
$f$
decreases continuously to reach zero
at 33.375$^\circ$C [Fig.~\ref{fig3}(a)],  while the
initial decay continues to about 0.3~s.
Thus, within this range of tempertures, for which the interparticle
attractions are increasing [Fig. \ref{fig2}(a)],
remarkably,
%it is remarkable that
the initially-arrested
RG melts into a
dynamic liquid.
The observed evolution of $g_1$ is distinct from what occurs when a hard sphere
suspension undergoes the liquid-glass transition as a function of
volume fraction. In that case, a plateau with a non-zero, $Q$-dependent
value
starts to emerge within the liquid phase, and,
with further increase in the volume fraction,
the plateau extends to longer times, showing only a small
increase in its value  through the glass
transition \cite{van_megan:91,gotze:91}.

On warming beyond 33.375$^\circ$C [Fig.~\ref{fig3}(b)],
the relaxation of $g_1$ becomes
progressively slower with increasing
temperature. On reaching 33.485$^\circ$C,
for times from  about 0.1 s to about 100 seconds,
$g_1$  appears nearly linear in the semi-log plot of Fig. \ref{fig3}(b),
indicating the onset of nearly logarithmic relaxations.
On warming still further, the logarithmic behavior
continues, but with a slope that decreases with increasing temperature, so
that by 33.600$^\circ$C, $g_1$  barely decreases to 0.95 after 300 seconds,
corresponding to an almost completely arrested state.
As noted previously, logarithmic  behavior of $g_1$
is expected for the liquid near an AG.

To quantify these observations,
for temperatures from 33.000$^\circ$C to 33.480$^\circ$C,
we have fitted $g_1$ using
an empirical, modified-stretched-exponential form:
$g_1 = \left [ f + (1-f)e^{-(t/t_0)^\alpha} \right ]
e^{-(t/t_2)^{3/2}}$.
To limit the number of parameters varied in the fits,
the stretching exponent ($\alpha$)
was fixed to $0.6$, leaving $f$, $t_0$ and $t_2$ as fitting parameters.
From 33.485$^\circ$C to 33.600$^\circ$C,
we have instead  modeled $g_1$ via an alternate empirical form:
$g_1= \left [ e^{-t/t_1}+(1-e^{-t/t_1})[1-S \log ( {1+t/t_1})
]
\right ]
 e^{-(t/t_2)^{3/2}}$,
which manifests logarithmic relaxations for $t_2 \gg t \gg t_1$.
On the semi-log axis of Fig.~\ref{fig3}(b),
$S$ is the slope of the logarithmic $g_1$.
In this case, the fitting parameters are $S$ and $t_2$, while
the onset of the logarithmic behavior ($t_1$) was fixed at 0.02~s.
The solid lines in Figs. \ref{fig3}(a) and \ref{fig3}(b)
correspond to the best fits to these
models,
providing a satisfactory account of the measured $g_1$ at all temperatures,
except at 33.480$^\circ$C, where $g_1$ is evolving from one
form to the other.
In both equations, the compressed exponential factor \cite{band:04}, involving
$t_2$, accounts for the behavior at long times
($t_2 \geq 200$~s)
and will not be discussed
further in this paper.

\begin{figure}
{\includegraphics[width=2.7in,keepaspectratio=true]{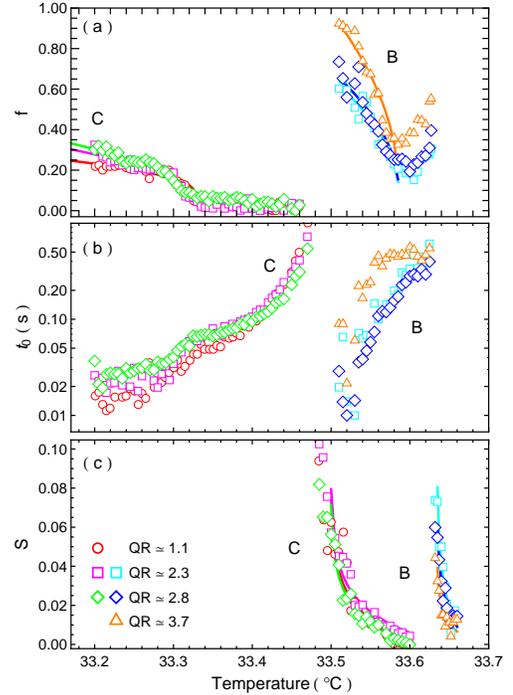}}
\caption{
(a) $f$, (b) $t_0$, and (c) $S$
{\em vs.} temperature for Run B
at $QR = 2.28$ (cyan squares), $2.86$ (blue diamonds) and $3.70$
(orange triangles),
and for Run C at $QR = 1.13$ (red circles), $2.34$ (magenta squares),
and $2.82$ (green diamonds).
%The peak of $S(Q)$ occurs at $QR=3.6$ to 3.7, depending on temperature.
The solid lines are guides-to-the-eye.
}
\label{fig4}
\end{figure}

Fig. \ref{fig4}(a) shows the best-fit values of
$f$
plotted versus temperature,  obtained during run B
at  $QR = 2.28$, 2.86, and 3.70
and during Run C at  $QR = 1.13$, 2.34, and 2.82.
As may be seen, $f$ takes a $Q$-dependent value
with systematically larger values
near the peak of the static structure factor
($QR \simeq 3.7$). This is
similar to what is observed for
hard-sphere glasses. However, here,
$f$ goes continuously to zero with increasing temperature.
This behavior stands in contrast to what occurs at
the melting of a hard-sphere glass with decreasing
density, where  the non-ergodicity parameter
goes to zero discontinuously.
MCT also predicts a  discontinuous decrease in the non-ergodicity
at the RG-to-liquid transition
in the presence of attractive interactions
%The MCT equations that apply to hard spheres exhibit a fold bifurcation.
%Consequently, the hard-sphere liquid-to-glass transition
%involves a discontinuous change in
%$f$ versus control parameter.
%By contrast, the MCT equations for attractive
%hard spheres exhibit cusp bifurcations.
%It follows that the transition from the liquid to
%the repulsive glass is, in this case, predicted to involve a
%continuous decrease of  $f$ to zero
\cite{PhysRevE.59.5706,PhysRevE.59.R1347,DawsonPRE2001}.
Thus, this aspect of our measurements presents a
contradiction to MCT.
It may be noted, though, that certain MCT models
\cite{PhysRevE.66.011405} do permit a glass
transition involving continuous increase in
$f$
as we observe in Run C.
For Run B, density correlations relax very slowly,
so that $g_1$ does not decrease to zero for times less than 200~s
at any temperature.
In fact, $g_1$ relaxes progressively more rapidly
for temperatures increasing to 33.55$^\circ$C, but relaxes
progressively more slowly for temperatures
increasing beyond 33.55$^\circ$C.
This leads to an apparent minimum in the fitted value of $f$
at 33.55$^\circ$C, before $g_1$ crosses over
to the logarithmic form for 33.64$^\circ$C and above.
We ascribe the difference in behavior between Run B and Run C
to the closer proximity of a glass-to-glass transition in Run B.

The best-fit values of the relaxation time ($t_0$)
are shown in Fig. \ref{fig4}(b) plotted on a logarithmic scale
versus temperature.
Evidently, $t_0$ depends strongly on temperature.
For both Runs B and C, $t_0$
increases rapidly with temperature from on the order of 0.02~s to
reach about 0.5~s, before the relaxations become logarithmic.
$t_0$ depends  relatively
weakly on $Q$, although in Run B
it is somewhat larger at the structure
factor peak ($QR=3.7$) than elsewhere.
For Run C, it was not possible to determine $t_0$ at the structure
factor peak because
of the presence of Bragg peaks from small coexisting crystallites.
In comparison, a $Q$-independent time-scale, that increases rapidly
with density, is a hallmark of
relaxations near the glass transition of hard spheres.

The best-fit values of $S$  within the logarithmic regime are summarized in
Fig. \ref{fig4}(c) versus temperature.
$S$ is only weakly $Q$-dependent.
Interestingly, near the temperature at which it vanishes,
$S$ appears to decrease approximately linearly to zero
with increasing temperature.
Although expressions for $S$
are available within the context of MCT for small displacements
in $u-\phi$ phase space from
the so-called $A_3$
singularity point \cite{PhysRevE.66.011405},
it is unclear how these predictions relate to
the behavior seen in Fig.~\ref{fig4}(c).

Our results for the phase behavior of silica particles
in WL in the volume fraction-interaction
strength ($\phi-u$) plane are summarized in
Fig. \ref{fig2}(b), where diamonds correspond to Run A, triangles
to Run B, and circles to Run C. Each point is plotted
with $u$ determined by fitting the SAXS lineshape and
with $\phi$ determined by the corresponding x-ray transmission.
Red symbols correspond to the RG,
defined
for the purposes of the figure
by the criterion that $g_1 > 0.17$ at
$t = 10$~s. Green symbols correspond to the AG, defined
similarly
via $g_1 > 0.81$ at $t = 10~s$. Blue symbols correspond to the liquid.
In the context of Refs.
\onlinecite{PhysRevE.59.5706,PhysRevE.59.R1347,DawsonPRE2001,PhysRevE.66.041402},
it is
natural to add the
schematic phase boundaries shown as solid lines in the figure.
The dashed line in the figure  separates the region of
logarithmic relaxations
above
the line from the region of stretched exponential relaxations below.
Overall, there is
good agreement between our experimental
phase diagram and that predicted.
In addition, this study reveals a number of previously unremarked aspects
including: (1)
A continuous decrease in the non-ergodicity parameter with attraction
strength through attraction-driven melting,
(2) the existence of a weakly $Q$-dependent time-scale that increases through
attraction-driven melting, and (3) the continuous
decrease of $S$ for temperatures increasing towards the liquid-to-attractive
glass transition.
%Beyond MCT, it is interesting to ask how
%the behavior presented here
%fits with alternate views of the
%glass transition \cite{LiuNature1998,TrappeNature2001,Geissler2005}?
We thank
T. Chiba, E. R.  Dufresne, R. L. Leheny, C. O'Hern,
S. Sanis, M. Spannuth, M. Trias,
and J. Wettlaufer for discussions, a referee for key clarifications, and
the NSF for support via
DMR 0453856.
The APS
is supported by the
DOE.

\end{document}